\documentclass[12pt]{article}
\usepackage{graphicx}
\usepackage{lineno}

\newcommand\pubnumber{}
\newcommand\pubdate{\today}

\def\institute{Technische Universit\"at Dortmund\\
Otto-Hahn Strasse 4a, 44227 Dortmund, Germany}

\def\Title#1{\begin{center} {\Large #1 } \end{center}}
\def\Author#1{\begin{center}{ \sc #1} \end{center}}
\def\Address#1{\begin{center}{ \it #1} \end{center}}

\newcommand\pubblock{\rightline{\begin{tabular}{l} \pubnumber\\
         \pubdate  \end{tabular}}}
\newenvironment{Abstract}{\begin{quotation}  }{\end{quotation}}
\newenvironment{Presented}{\begin{quotation} \begin{center} 
             PRESENTED AT\end{center}\bigskip 
      \begin{center}\begin{large}}{\end{large}\end{center} \end{quotation}}
\def\Acknowledgements{\bigskip  \bigskip \begin{center} \begin{large}
             \bf ACKNOWLEDGEMENTS \end{large}\end{center}}

\def\beq{\begin{equation}}
\def\eeq#1{\label{#1}\end{equation}}
\def\eeqn{\end{equation}}

\def\beqa{\begin{eqnarray}}
\def\eeqa#1{\label{#1}\end{eqnarray}}
\def\eeqan{\end{eqnarray}}

\let\bar=\overbar

\def\Dslash{\not{\hbox{\kern-4pt $D$}}}
\def\dslash{\not{\hbox{\kern-2pt $\del$}}}

\def\msb{{\bar{\ssstyle M \kern -1pt S}}}

\mathchardef\mhyphen="2D

\begin{document}
\begin{titlepage}
\pubblock

\vfill
\Title{Search for $t\bar{t}H$ and $tH$ production with $H\to \gamma \gamma$ at $\sqrt{s}$~=~13~TeV with the ATLAS experiment}
\vfill
\Author{ Isabel  Nitsche \\ On behalf of the ATLAS Collaboration}
\Address{\institute}
\vfill
\begin{Abstract}
The measurement of the Yukawa coupling of the top quark is one important goal of particle physics after the discovery of the Higgs boson. There are two processes that offer the possibility to directly measure the Yukawa coupling of the top quark at the LHC. These are the production of a top-antitop-quark pair in association with a Higgs boson via the strong interaction ($t\bar{t}H$) and the electroweak production of a single top quark and a Higgs boson ($tH$). In the Standard Model, the cross section for $tH$ production is much smaller than the cross section for $t\bar{t}H$ production as there is a destructive interference between diagrams where the Higgs boson is radiated off a top quark and diagrams where it is radiated off a \textit{W} boson. Despite the reduced cross section, this interference provides additional sensitivity to the sign of the Yukawa coupling. 

The considered Higgs boson decay into photons is promising due to the good energy resolution of the photons and the relatively small backgrounds. But it is also challenging due to the small branching ratio. 

The analysis strategy and the latest results on the search for $t\bar{t}H/tH$ production in the diphoton channel at the ATLAS experiment are presented.
\end{Abstract}
\vfill
\begin{Presented}
$10^{th}$ International Workshop on Top Quark Physics\\
Braga, Portugal,  September 17--22, 2017
\end{Presented}
\vfill
\end{titlepage}
\def\thefootnote{\fnsymbol{footnote}}
\setcounter{footnote}{0}

\section{Introduction}
\begin{figure}[htb]
\centering
\includegraphics[height=2.5in]{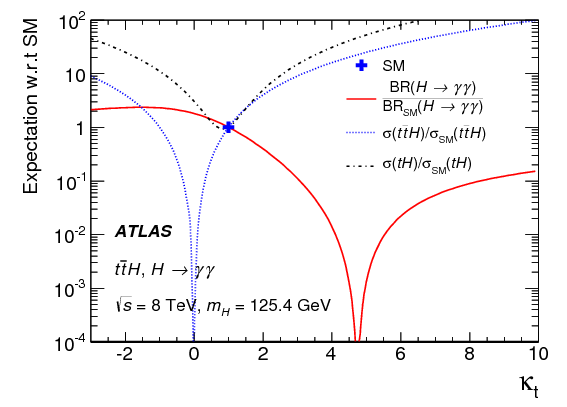}
\caption{The ratio of $\sigma(tH)$, $\sigma(t\bar{t}H)$ and $B(H \to \gamma \gamma)$  to the SM expectation for different values of $\kappa_t$ at $\sqrt{s}$ = 8 TeV \cite{8tev}.}
\label{fig:kappat}
\end{figure}

Due to the large mass of the top quark, it is expected to couple strongly to the Higgs boson as predicted by the Standard Model (SM). Therefore, the measurement of the Yukawa coupling of the top quark $Y_t$ is an important test of the SM. This coupling can be measured at the LHC either indirectly through processes involving top quark loops or directly in processes with real top quarks. The associated production of the Higgs boson and a top quark pair ($t\bar{t}H$) or a single top quark ($tH$) allows for the direct measurement of $Y_t$. A scale factor $\kappa_t$ can be introduced to describe deviations from the SM expectation: $Y_t$ = $\kappa_t \times Y_t^{SM}$. In $tH$ production, the Higgs boson can be radiated either off the top quark or the \textit{W} boson. There is destructive interference of these two processes in the SM leading to a small cross section of $tH$ production. But this interference in $tH$ production also provides sensitivity to negative values of $\kappa_t$. For values $\kappa_t$ $\neq$ 1, the cross section $\sigma(tH)$ would strongly increase, as shown in Figure \ref{fig:kappat} for $\sqrt{s}$ = 8 TeV. A similar behaviour is expected at $\sqrt{s}$ = 13 TeV. 

The measurement of $t\bar{t}H$ and $tH$ production  at $\sqrt{s}$ = 13~TeV performed in the $H \to \gamma \gamma$ channel with 36.1 $\mathrm{fb}^{-1}$ \cite{Conf} at the ATLAS experiment \cite{atlas} is presented here.
In this measurement, a global fit is performed to categories targeting several Higgs production processes in the $H \to \gamma \gamma$ channel: gluon fusion ($gg \to H$), vector-boson fusion (VBF), the production of a Higgs boson in association with a vector boson ($VH$) and top-associated production ($t\bar{t}H$+$tH$).   

\section{Analysis strategy}

The analysis strategy in the $H \to \gamma \gamma$ decay channel is based on the good photon energy resolution of the electromagnetic calorimeter of the ATLAS detector. The diphoton invariant mass  distribution $m_{\gamma \gamma}$ has a narrow signal peak at the Higgs boson mass of 125.09 GeV with a width of approximately 1.5~GeV. Only events with two  photons fulfilling certain requirements are considered. Based on this photon selection, categories targeting different final states are defined.

There are two different kinds of background that need to be considered: the Higgs boson background and the continuum background. The continuum background consists of events with two photons that do not originate from a Higgs boson decay. It has a falling $m_{\gamma \gamma}$ distribution. While the contributions of the Higgs boson backgrounds can be studied in Monte Carlo simulations, the composition of the continuum background is not well known and therefore it is estimated by performing a fit to data sidebands with an analytic function. Different functions with a varying number of degrees of freedom are taken into account for each category. The functional form is chosen by performing a signal+background fit to a background-only template. Finally, the function minimizing the bias in the extracted signal yield is chosen. For the $t\bar{t}H$ and $tH$ categories, an exponential or a power law function with only one degree of freedom is chosen.

In addition, a signal model is defined for each category. The signal is modelled by a double-sided crystal ball function whose parameters are determined from a fit to Monte Carlo simulations. Finally, an unbinned likelihood fit to $m_{\gamma \gamma}$ is performed for each category to estimate the observed signal yield.

\section{Event selection}


After the diphoton preselection, nine categories targeting the $t\bar{t}H$ and $tH$ final states are defined. Final states with and without leptons are considered separately. A summary of the categorization is shown in Table \ref{tab:ttH_tH_cat}. The categories are filled in the order shown in Table \ref{tab:ttH_tH_cat} to ensure their orthogonality.

Due to the smaller branching ratio of the leptonic decays, the categories requiring leptons in the final state are defined first. A cut-based approach is used as it provides a sufficient background rejection. For all categories, at least one lepton and at least one $b$-tagged jet is required. Two categories are targeting mainly $tH$ production, by exploiting the presence of a forward jet in the $tH$ $t$-channel production and using a requirement on the maximum number of jets. The third category is targeting mainly $t\bar{t}H$ and allows a higher number of jets and a second lepton. For dilepton events, a \textit{Z} veto is applied.

The selection of events without leptons in the final state is more challenging due to the larger background. Therefore, an approach based on a boosted decision tree (BDT) is used here. The BDT is trained in order to separate $t\bar{t}H$ from $gg \to H$ and multijet background. A preselection is applied, requiring zero leptons, at least three jets and at least one $b$-tagged jet. Five input variables are used to train the BDT: the scalar sum of the $p_\mathrm{T}$ of the jets $H_\mathrm{T}$, the mass of all jets $m_{\mathrm{all jets}}$, the total number of jets $N_{\mathrm{jets}}$, the number of central jets $N_{\mathrm{jets}}^{\mathrm{cen}}$ ($|\eta|$ $<$ 2.5) and the number of $b$-tagged jets $N_{b\mhyphen\mathrm{tag}}$. Four categories are defined by cutting on the BDT output discriminant. In addition, two cut-based $tH$ categories are defined, requiring exactly four central jets and either one or two $b$-tagged jets.     

\begin{table}[t]
\begin{center}
\begin{tabular}{l|l}
\hline\hline
Category & Selection \\
\hline\hline
$tH$ lep 0fwd & $N_{\mathrm{lep}}$ = 1, $N_{\mathrm{jets}}^{\mathrm{cen}}$ $\le$ 3, $N_{b\mhyphen\mathrm{tag}}$ $\ge$ 1, $N_{\mathrm{jets}}^{\mathrm{fwd}}$ = 0 ($p_\mathrm{T}^{\mathrm{jet}}$ $>$ 25 GeV) \\
$tH$ lep 1fwd & $N_{\mathrm{lep}}$ = 1, $N_{\mathrm{jets}}^{\mathrm{cen}}$ $\le$ 4, $N_{b\mhyphen\mathrm{tag}}$ $\ge$ 1, $N_{\mathrm{jets}}^{\mathrm{fwd}}$ $\ge$ 1 ($p_\mathrm{T}^{\mathrm{jet}}$ $>$ 25 GeV) \\
$t\bar{t}H$ lep & $N_{\mathrm{lep}}$ $\ge$ 1, $N_{\mathrm{jets}}^{\mathrm{cen}}$ $\ge$ 2, $N_{b\mhyphen\mathrm{tag}}$ $\ge$ 1, $Z_{\ell \ell}$ veto  ($p_\mathrm{T}^{\mathrm{jet}}$ $>$ 25 GeV) \\ 
$t\bar{t}H$ had BDT1 & $N_{\mathrm{lep}}$ = 0, $N_{\mathrm{jets}} \geq$ 3,     $N_{b\mhyphen\mathrm{tag}} \geq$ 1, $BDT_{ttH}$ $>$ 0.92  \\
$t\bar{t}H$ had BDT2 & $N_{\mathrm{lep}}$ = 0, $N_{\mathrm{jets}} \geq$ 3,     $N_{b\mhyphen\mathrm{tag}} \geq$ 1, 0.83 $<$ $BDT_{ttH}$ $<$ 0.92  \\
$t\bar{t}H$ had BDT3 & $N_{\mathrm{lep}}$ = 0, $N_{\mathrm{jets}} \geq$ 3,     $N_{b\mhyphen\mathrm{tag}} \geq$ 1, 0.79 $<$ $BDT_{ttH}$ $<$ 0.83 \\
$t\bar{t}H$ had BDT4 & $N_{\mathrm{lep}}$ = 0, $N_{\mathrm{jets}} \geq$ 3,     $N_{b\mhyphen\mathrm{tag}} \geq$ 1, 0.52 $<$ $BDT_{ttH}$ $<$ 0.79 \\ 
$tH$ had 4j1b & $N_{\mathrm{lep}}$ = 0, $N_{\mathrm{jets}}^{\mathrm{cen}}$ = 4,    $N_{b\mhyphen\mathrm{tag}}$ = 1 ($p_\mathrm{T}^{\mathrm{jet}}$ $>$ 25 GeV) \\
$tH$ had 4j2b & $N_{\mathrm{lep}}$ = 0, $N_{\mathrm{jets}}^{\mathrm{cen}}$ = 4,    $N_{b\mhyphen\mathrm{tag}}$ $\geq$ 2 ($p_\mathrm{T}^{\mathrm{jet}}$ $>$ 25 GeV) \\ 
\hline\hline
\end{tabular}
\caption{Summary of the nine categories targeting leptonic and hadronic $t\bar{t}H$ and $tH$ final states \cite{Conf}.}
\label{tab:ttH_tH_cat}
\end{center}
\end{table}

\section{Results}

\begin{figure}[h]
\centering
\includegraphics[height=2.8in]{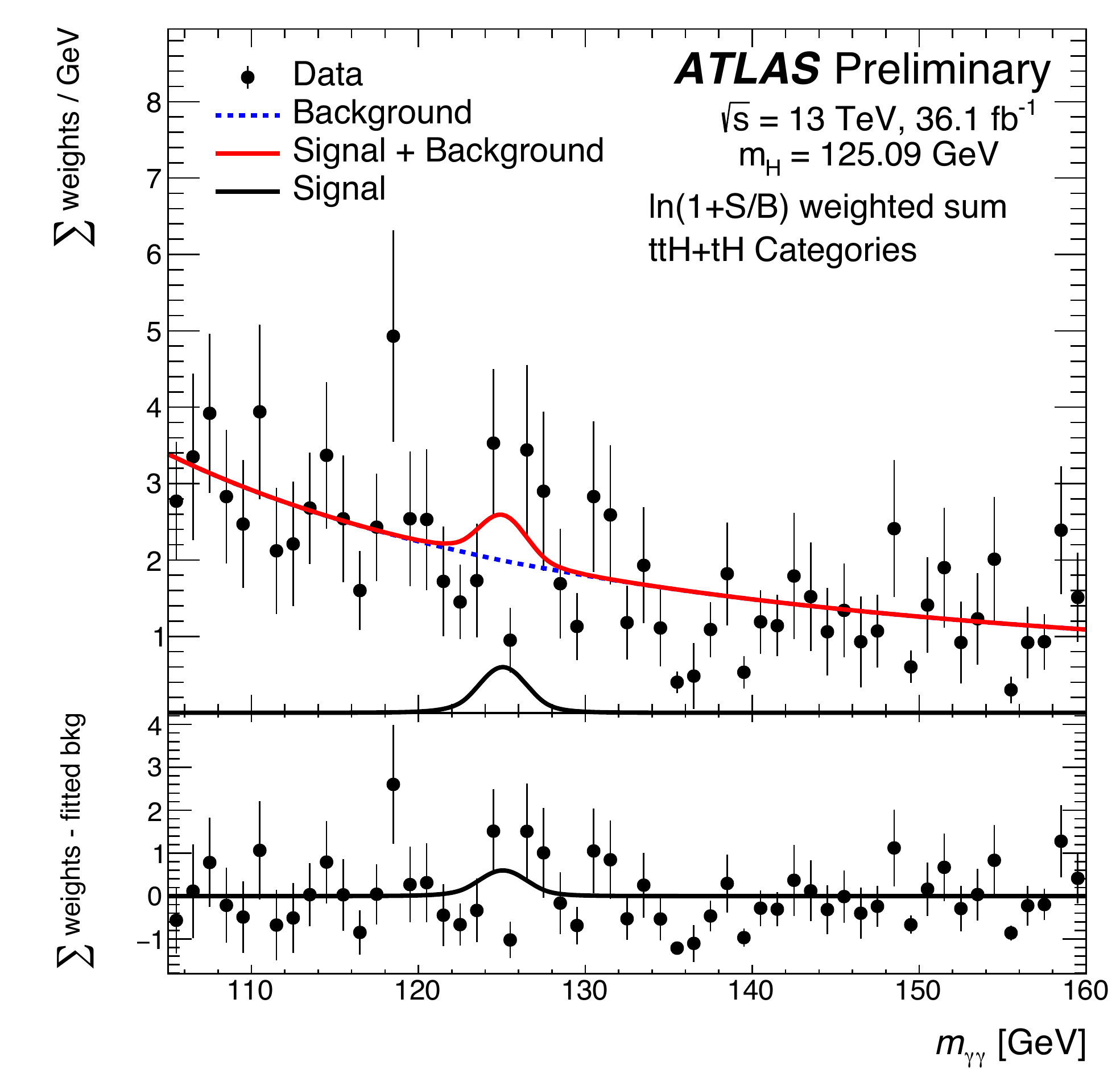}
\caption{The diphoton invariant mass spectrum of the $t\bar{t}H$ and $tH$ categories \cite{Conf}.}
\label{fig:inv_mass}
\end{figure}

A measurement of the production cross section and the signal strength is performed for top-associated Higgs boson production 
in the $H \to \gamma \gamma$ channel. The diphoton invariant mass spectrum of the $t\bar{t}H/tH$ categories is shown in Figure~\ref{fig:inv_mass}. 
The measurement of the production cross section for $|y_H|$  $<$ 2.5 results in
$\sigma_{\mathrm{top}} \times B(H \to \gamma \gamma)$ = 0.7\,$^{+0.9}_{-0.7}$~fb, while the SM predicts $\sigma_{\mathrm{top}} \times B(H \to \gamma \gamma)$ = 1.3\,$^{+0.9}_{-0.8}$~fb.
This corresponds to a measured signal strength of 
$\mu_{\mathrm{top}}$\,=\,0.5\,$_{-0.5}^{+0.6}$\,(stat.)\,$_{-0.1}^{+0.1}$\,(exp.)\,$_{-0.0}^{+0.1}$\,(theory), leading to an observed (expected) significance of 1.0 $\sigma$ (1.8 $\sigma$). As these results show no evidence for top-associated production, a  95\% CL limit of 1.7$\mu$ on the signal strength is derived. Several sources of systematic uncertainties are considered, which can be grouped into uncertainties due to the choice of signal and background model, experimental effects and theory uncertainties. For the total uncertainty, these systematic uncertainties only have a small impact, because the statistical uncertainty dominates the total uncertainty.

\section{Conclusion}

The measurement of the production cross section and the signal strength of top-associated production in the $H \to \gamma \gamma$ channel at $\sqrt{s}$ = 13 TeV and 36.1 $\mathrm{fb}^{-1}$ shows no deviations from the SM. As no evidence was found for this process though, a limit on the signal strength was set. New analysis techniques have been used by ATLAS in this channel. A BDT is used in the hadronic channel in order to separate $t\bar{t}H$ from multijet and $gg \to H$ background.
In addition, dedicated categories enriched in $tH$ are defined in order to be sensitive to negative values of $\kappa_t$.

\Acknowledgements
The author acknowledges the financial support by the Federal Ministry of Education and Research of Germany in the framework of ATLAS (FSP 103).

\end{document}